\documentclass[aps,amsmath,amssymb,showpacs,nofootinbib]{revtex4}
\usepackage{graphicx}
\usepackage{amssymb}

\begin{document}
\title{Phase diagram of two-flavor quark matter: 
Gluonic phase at nonzero temperature}
\author{O. Kiriyama}
\email{kiriyama@th.physik.uni-frankfurt.de}
\affiliation{Institut f\"ur Theoretische Physik, 
J.W.\ Goethe-Universit\"at, D-60438 Frankfurt am Main, Germany\\
and Research Center for Nuclear Physics, 
Osaka University, Ibaraki 567-0047, Japan}

\begin{abstract}
The phase structure of neutral two-flavor quark matter 
at nonzero temperature is studied. 
Our analysis is performed within the framework 
of a gauged Nambu--Jona-Lasinio model and the mean-field approximation. 
We compute the free energy 
of the gluonic phase (gluonic cylindrical phase II) 
in a self-consistent manner and investigate the phase transition 
from the gluonic phase to the 2SC/g2SC/NQ phases. 
We briefly consider the phase diagram in the plane of 
coupling strength versus temperature 
and discuss the mixed phase consisting of the normal quark and 2SC phases.
\end{abstract}
\date{\today}
\pacs{12.38.-t, 11.30.Qc, 26.60.+c}
\maketitle

\section{Introduction}
The properties of cold and dense quark matter 
are of great interest in astrophysics and cosmology. 
In particular, at moderate densities of relevance for 
the interior of compact stars, quark matter is a color superconductor 
and has a rich phase structure 
with important implications for compact star physics 
\cite{CSC1,CSC2,CSC3,CSC4,CSC5,CSC6,CSC7,CSC8,CSC9}.

Bulk matter in the interior of compact stars 
should be color and electrically neutral and be in $\beta$-equilibrium. 
In the two-flavor case, these conditions separate 
the Fermi momenta of up and down quarks and, 
as a consequence, the ordinary BCS state (2SC) is not always 
energetically favored over other unconventional states. 
The possibilities include crystalline color superconductivity 
and gapless color superconductivity (g2SC) 
\cite{ABR,BR,Shovkovy2003,Shovkovy2003b}. 
However, the 2SC/g2SC phases suffer from a chromomagnetic instability, 
indicated by imaginary Meissner masses of some gluons 
\cite{Huang2004,Huang2004b}.
The instability related to gluons of color 4--7 
occurs when the ratio of the gap 
over the chemical potential mismatch, $\Delta/\delta\mu$, 
decreases below a value $\sqrt{2}$. 
Resolving the chromomagnetic instability 
and clarifying the nature of true ground state of dense quark matter 
are central issues in the study of color superconductivity 
\cite{Giannakis2004,Giannakis2005,Giannakis2005b,RedRup,
Neumann,SHH,Huang2005,Hong2005,Gorbar2005,Gorbar2005a,
Gorbar2005b,Fukush2006,GHMS2006,Hashimoto2006,
KRS2006,Kiri2006,Kiri2006b,Iida2006,HJZ,Gatto2007,Hashimoto2007,Ferrer}. 
(For a three-flavor case, see Refs. \cite{CGMNR2005,Fukush2005,
supercurrent,supercurrent2,CNRG2006,Rajagopal2006,ZK2007}.)

\begin{figure}
\includegraphics[width=0.45\textwidth]{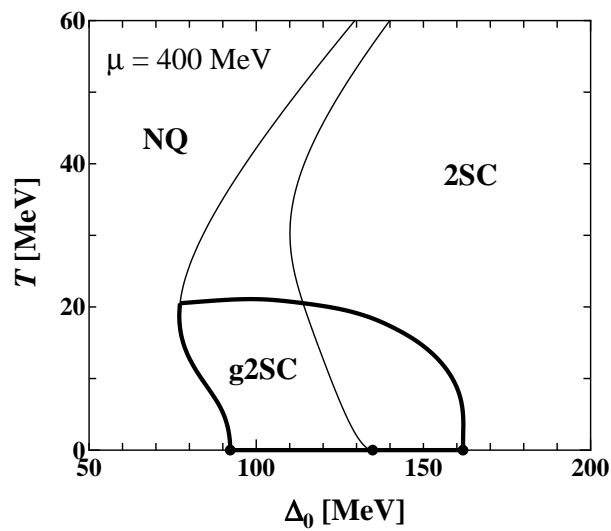}
\caption{The phase diagram of electrically neutral 
two-flavor quark matter in the plane of 
$\Delta_0$ and $T$. At $T=0$, the g2SC phase 
exists in the window 
$92~{\rm MeV} < \Delta_0 < 134~{\rm MeV}$ 
and the 2SC window is given by 
$\Delta_0 > 134~{\rm MeV}$. 
The unstable region for gluons 4--7 is depicted 
by the region enclosed by the thick solid line. 
The g2SC phase and a part of the 2SC phase 
($92~{\rm MeV} < \Delta_0 < 162~{\rm MeV}$) suffer from 
the chromomagnetic instability at $T=0$. 
The quark chemical potential 
is taken to be $\mu=400~{\rm MeV}$.}
\label{Figure1}
\end{figure}

As an example, in Fig. \ref{Figure1}, we plot the phase diagram 
of neutral two-flavor quark matter 
in the plane of the 2SC gap at $\delta\mu=0$ ($\Delta_0$) 
and temperature ($T$) \cite{Kiri2006,Kiri2006b}. 
(The parameter $\Delta_0$ is essentially the diquark coupling strength.) 
In order to obtain this diagram, we employed 
a gauged Nambu--Jona-Lasinio (NJL) model, 
which is the very same model that we shall use in this paper. 
Furthermore, we neglected the color chemical potential. 
The 2SC/g2SC phases 
and the unpaired normal quark (NQ) phase were included 
in the analysis. 
The quark chemical potential was taken to be $\mu=400~{\rm MeV}$, 
which is a value typical for the cores of compact stars. 
The region enclosed by the thick solid line is unstable 
(because gluons of color 4--7 have tachyonic Meissner masses there) 
and, therefore, should be replaced by 
other chromomagnetically stable phases, 
for instance, gluonic phases \cite{Gorbar2005,Gorbar2005a}. 
(For more detailed discussions of the gluonic phases, see Refs. \cite{VC,VC2}.) 
Note, however, that we did not consider the global structure of a free energy 
in extracting the unstable region, but only the tendency 
toward the vector condensation 
$\langle \vec{A^6} \rangle$ in the 2SC/g2SC phases. 
A self-consistent analysis of the gluonic phases at $T=0$ 
has recently been done by Hashimoto and Miransky \cite{Hashimoto2007} 
and they found that the gluonic phase 
(strictly speaking, the gluonic cylindrical phase II) 
exists in the window $65~{\rm MeV} < \Delta_0 < 160~{\rm MeV}$ 
and is energetically more favored than the 2SC/g2SC/NQ phases 
in this whole window. 

In this paper, we study the gluonic cylindrical phase II 
at nonzero temperature and revisit 
the phase diagram shown in Fig. \ref{Figure1}, 
computing the free energy of the gluonic phase 
in a self-consistent manner. 
The result would be useful for the phase diagram of QCD, 
and the compact star phenomenology as well.

\section{Model}
In order to study the gluonic phase, 
we use the gauged NJL model 
with massless up and down quarks:
\begin{eqnarray}
{\cal L}=\bar{\psi}(iD\hspace{-7pt}/+\hat{\mu}\gamma^0)\psi
+G_D\left(\bar{\psi}i\gamma_5\varepsilon\epsilon^bC\bar{\psi}^T\right)
\left(\psi Ci\gamma_5\varepsilon\epsilon^b\psi\right)
-\frac{1}{4}F_{\mu\nu}^{a}F^{a\mu\nu},
\end{eqnarray}
where the quark field $\psi$ carries flavor ($i,j=1,\ldots N_f$ 
with $N_f=2$) and color ($\alpha,\beta=1,\ldots N_c$ with $N_c=3$) 
indices, $C$ is the charge conjugation matrix; 
$(\varepsilon)^{ik}=\varepsilon^{ik}$ and 
$(\epsilon^b)^{\alpha\beta}=\epsilon^{b\alpha\beta}$ 
are the antisymmetric tensors in flavor and color spaces, 
respectively. The diquark coupling strength 
in the scalar ($J^P=0^+$) color-antitriplet channel 
is denoted by $G_D$. The covariant derivative and the field 
strength tensor are defined as
\begin{subequations}
\begin{eqnarray}
D_{\mu} &=& \partial_{\mu}-igA_{\mu}^{a}T^{a},\\
F_{\mu\nu}^{a} &=& \partial_{\mu}A_{\nu}^{a}-\partial_{\nu}A_{\mu}^{a}
+gf^{abc}A_{\mu}^{b}A_{\nu}^{c}.
\end{eqnarray}
\end{subequations}
To evaluate loop diagrams we use a three-momentum cutoff 
$\Lambda=653.3$ MeV throughout this paper. 
In NJL-type models without dynamic gauge fields, 
one has to introduce color 
and electric chemical potentials ($\mu_8$ and $\mu_e$) by hand \cite{BubSho} 
to ensure color- and electric-charge neutrality. 
In $\beta$-equilibrated neutral two-flavor quark matter, 
the elements of the diagonal matrix of 
quark chemical potentials $\hat{\mu}$ are given by
\begin{eqnarray}
&&\mu_{ur}=\mu_{ug}=\bar{\mu}-\delta\mu,\nonumber\\
&&\mu_{dr}=\mu_{dg}=\bar{\mu}+\delta\mu,\nonumber\\
&&\mu_{ub}=\bar{\mu}-\delta\mu-\mu_8,\nonumber\\
&&\mu_{db}=\bar{\mu}+\delta\mu-\mu_8,
\end{eqnarray}
with
\begin{eqnarray}
\bar{\mu}=\mu-\frac{\delta\mu}{3}+\frac{\mu_8}{3},
~\delta\mu=\frac{\mu_e}{2}.
\end{eqnarray}

In Nambu-Gor'kov space, the inverse full quark propagator 
$S^{-1}(p)$ is written as
\begin{eqnarray}
S^{-1}(p)=\left(
\begin{array}{cc}
(S_0^+)^{-1} & \Phi^- \\
\Phi^+ & (S_0^-)^{-1} 
\end{array}
\right),\label{eqn:propagator}
\end{eqnarray}
with
\begin{subequations}
\begin{eqnarray}
&&(S_0^+)^{-1}=\gamma^{\mu}p_{\mu}+(\bar{\mu}-\delta\mu\tau^3)\gamma^0
+g\gamma^{\mu}A_{\mu}^{a}T^{a},\\
&&(S_0^-)^{-1}=\gamma^{\mu}p_{\mu}-(\bar{\mu}-\delta\mu\tau^3)\gamma^0
-g\gamma^{\mu}A_{\mu}^{a}T^{aT},
\end{eqnarray}
\end{subequations}
and
\begin{eqnarray}
\Phi^- = -i\varepsilon\epsilon^b\gamma_5\Delta,~
\Phi^+ = -i\varepsilon\epsilon^b\gamma_5\Delta.
\end{eqnarray}
Here $\tau^3=\mbox{diag}(1,-1)$ is a matrix in flavor space. Following the 
usual convention, we have chosen the diquark condensate to point 
in the third direction in color space. 

For the gluonic cylindrical phase II, 
$B=\langle gA_z^6 \rangle$ is the most relevant condensate, 
because the chromomagnetic instability related to gluons 4--7 
corresponds to the tachyonic mode 
in the direction of $B$.\cite{Gorbar2005,Gorbar2005a,KRS2006} 
Besides $B$, we have to introduce a color chemical potential 
$\mu_3=\langle gA_0^3 \rangle$ to ensure color neutrality 
at $B \neq 0$. 
Taking into account these condensates, 
the free energy of the gluonic phase in the one-loop approximation 
is given by
\begin{eqnarray}
&&\varOmega(\Delta,\mu_e,\mu_8,B,\mu_3;\mu,T)\nonumber\\
&&\hspace{30pt}=-\frac{\mu_8^2B^2}{2g^2}
+\frac{\mu_3\mu_8B^2}{2g^2}-\frac{\mu_3^2B^2}{8g^2}\nonumber\\
&&\hspace{30pt}-\frac{1}{12\pi^2}
\left(\mu_e^4+2\pi^2T^2\mu_e^2+\frac{7\pi^4}{15}T^4\right)\nonumber\\
&&\hspace{30pt}+\frac{\Delta^2}{4G_D}
-\frac{1}{2}\sum_{a}\int\frac{d^3p}{(2\pi)^3}
\left[|\epsilon_a|+2T\ln(1+e^{-\beta|\epsilon_a|})\right],\label{eqn:ep1}
\end{eqnarray}
where $\beta=1/T$, the $\epsilon_a$'s are quasi-quark energies 
and the sum runs over all particle and anti-particle $\epsilon_a$'s. 
Here, we added tree-level contributions from gluons 
(first line on the r.h.s.),
\begin{eqnarray}
\varOmega_g^{{\rm (tree)}}=\frac{g^2}{4}f^{abc}f^{ade}
A_{\mu}^bA_{\nu}^cA^{d\mu}A^{e\nu},
\end{eqnarray}
and electrons (second line on the r.h.s.). 
Note also that the $\epsilon_a$'s depend on the vector condensates 
through the covariant derivatives 
in the quark propagator (\ref{eqn:propagator}). 
In what follows, we neglect the color chemical potentials 
$\mu_{3,8}$ and, consequently, 
the tree-level contributions of gluons. 
We have carefully checked that their effect on the free energy is negligible 
for realistic values of $\alpha_s \simeq 1$ 
(see also Ref. \cite{Hashimoto2007}). 

In this work, in order to remove the ultraviolet divergence 
in the Meissner screening masses we shall use the following subtraction
\begin{eqnarray}
\varOmega_R=\varOmega(\Delta,\mu_e,B;\mu,T)
-\varOmega(0,0,B;0,0).\label{eqn:ep2}
\end{eqnarray}
It is known that this free energy subtraction is not 
adequate to remove the cutoff dependence of the free energy at $T>0$. 
In fact, Eq. (\ref{eqn:ep2}) leads to 
positive Meissner screening masses 
in the normal phase at $T>0$ \cite{Kiri2006,Kiri2006b}. 
In this work we do not go into this problem because 
this unphysical behavior of the Meissner masses 
is nothing but a cutoff artifact 
and moreover is negligibly small at $\mu=400~{\rm MeV}$ 
and at the temperatures of interest (20 MeV at most). 

In order to find the neutral gluonic phase, 
we first solve a set of coupled equations 
(the gap equation and the electrical charge neutrality condition), 
\begin{eqnarray}
\frac{\partial\varOmega_R}{\partial\Delta}
=\frac{\partial\varOmega_R}{\partial\mu_e}=0,\label{eqn:eqs}
\end{eqnarray}
as a function of $B$ and, then, 
compute the free energy $\varOmega_R(B)$. 
Finally, the minimum of $\varOmega_R(B)$ determines 
the neutral gluonic phase. 
(In the following Figs. \ref{Figure3} and \ref{Figure4}, 
we plot the free energy evaluated along the solution 
of the coupled equations (\ref{eqn:eqs}).)

\section{Numerical Results}
\begin{figure}
\includegraphics[width=0.45\textwidth]{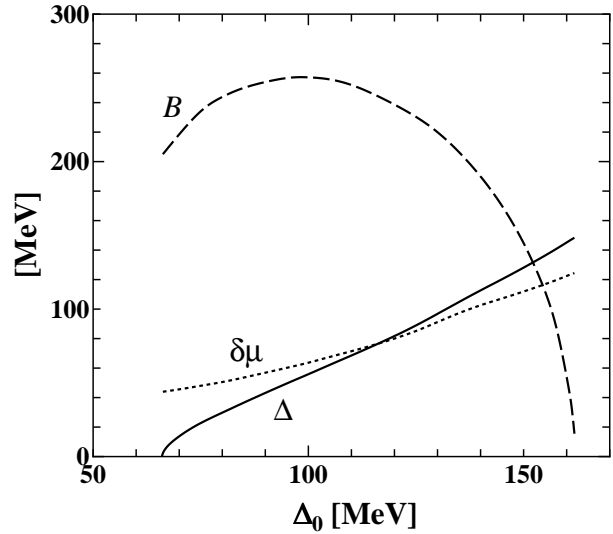}
\caption{The gap parameter $\Delta$ (solid line), 
the chemical potential mismatch $\delta\mu$ (dotted line) 
and the gluonic vector condensate $B$ (dashed line) versus $\Delta_0$ 
in the gluonic phase at $T=0$. 
The quark chemical potential is taken to be $\mu=400~{\rm MeV}$.}
\label{Figure2}
\end{figure}

Figure \ref{Figure2} shows $\Delta$, $\delta\mu$ and $B$ 
in the gluonic phase at $T=0$ as a function of $\Delta_0$. 
First, let us note that the results of Fig. \ref{Figure2} 
are in good agreement with those shown 
in Figs. 1, 2, and 3 of Ref. \cite{Hashimoto2007}, where the color chemical 
potentials $\mu_{3,8}$ were treated self-consistently. 
In Fig. \ref{Figure2} one can see that the gluonic phase exists in the window
\begin{eqnarray}
66~{\rm MeV} < \Delta_0 < 162~{\rm MeV}.
\end{eqnarray}
The gluonic phase is energetically favored over 
the 2SC/g2SC/NQ phases in this whole window 
(see also Fig. 5 of Ref. \cite{Hashimoto2007}). 
One also sees that the phase transition between the gluonic phase 
and the NQ (2SC) phase at $\Delta_0=66~{\rm MeV}$ ($162~{\rm MeV}$) 
is strongly (weakly) of first order. 

\begin{figure}
\includegraphics[width=0.45\textwidth]{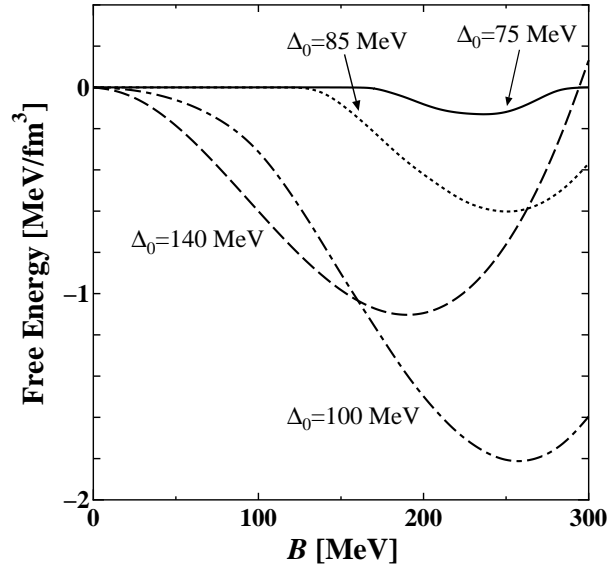}
\caption{The free energy $\varOmega_R(B)$ as a function of $B$ at $T=0$ 
for $\Delta_0=75~{\rm MeV}$ (solid line), 
$\Delta_0=85~{\rm MeV}$ (dotted line), 
$\Delta_0=100~{\rm MeV}$ (dot-dashed line), 
and $\Delta_0=140~{\rm MeV}$ (dashed line). 
Note that the free energy is measured with respect to 
the 2SC/g2SC/NQ phases at $B=0$. 
The results are plotted for $\mu=400~{\rm MeV}$.}
\label{Figure3}
\end{figure}

Now let us take a closer look at the free energy at $T=0$. 
Figure \ref{Figure3} shows the behavior of $\varOmega_R(B)$ 
measured with respect to the 2SC/g2SC/NQ phases at $B=0$. 
The results are plotted for $\mu=400~{\rm MeV}$ 
at several values of $\Delta_0$. 

In the weak coupling regime, $66~{\rm MeV} < \Delta_0 < 92~{\rm MeV}$, 
the chromomagnetic instability does not exist in the NQ phase 
(see Fig. \ref{Figure1}). 
We note that the curvature of $\varOmega_R(B)$ at $B=0$,
\begin{eqnarray}
m_M^2=\frac{d^2\varOmega_R(B)}{dB^2}\bigg{|}_{B=0},
\end{eqnarray}
can be regarded as the Meissner mass squared 
$\partial^2\varOmega_R/\partial B^2|_{B=0}$ in the 2SC/g2SC/NQ phases, 
since the solutions of Eq. (\ref{eqn:eqs}) 
satisfy $\Delta=\bar{\Delta}+{\cal O}(B^2)$ 
and $\mu_e=\bar{\mu}_e+{\cal O}(B^2)$ 
for small values of $B$, where $\bar{\Delta}$ and $\bar{\mu}_e$ 
denote their values at $B=0$ \cite{Gorbar2005,Gorbar2005a}. 
We found that $m_M^2$ is indeed zero 
in the weak coupling regime. 
In addition, we observed that, for small values of $B$, 
the system is in the ungapped ($\Delta=0$) phase 
and the free energy behaves like $\varOmega_R \sim {\cal O}(B^4)$. 
However, contrary to the result of Fig. \ref{Figure1}, 
the free energy has a global minimum 
at $B \neq 0$ and the gluonic phase is energetically 
favored over the NQ phase. 
For $92~{\rm MeV} < \Delta_0 \lesssim 162~{\rm MeV}$, 
one finds tachyonic modes at $B=0$ 
because the g2SC phase and a part of the 2SC phase suffer from 
the chromomagnetic instability and, therefore, are unstable 
against the formation of $B$. 
Consequently, the gluonic phase is realized in this region, as expected. 
For strong coupling, $\Delta_0 \gtrsim 162~{\rm MeV}$, 
the 2SC phase is chromomagnetically stable in this regime and 
the free energy has a global minimum at $B=0$, 
though it is not plotted in Fig. \ref{Figure3}.

\begin{figure}
\includegraphics[width=0.8\textwidth]{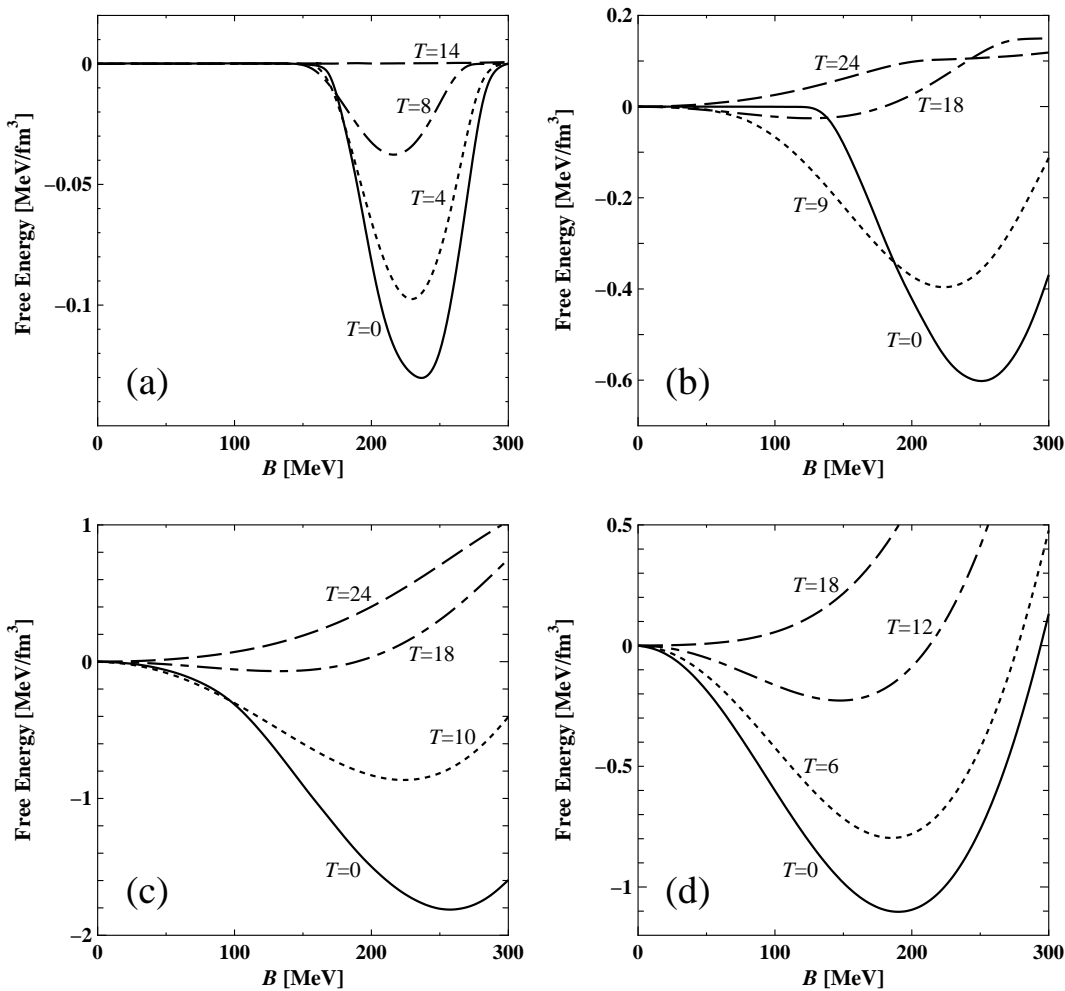}
\caption{The temperature dependence of the free energy 
(measured with respect to the 2SC/g2SC/NQ phases at $B=0$) 
as a function of $B$ 
for $\Delta_0=75~{\rm MeV}$ (a), 
for $\Delta_0=85~{\rm MeV}$ (b), 
for $\Delta_0=100~{\rm MeV}$ (c), 
and for $\Delta_0=140~{\rm MeV}$ (d). 
The results are plotted for $\mu=400~{\rm MeV}$ 
and the values of $T$ are given in MeV.}
\label{Figure4}
\end{figure}

We now turn to the free energy of the gluonic phase at $T > 0$. 
Figure \ref{Figure4}(a) display the temperature dependence 
of the free energy for $\Delta_0=75~{\rm MeV}$. 
As $T$ grows, the free-energy gain 
gets reduced, but the change of the vacuum expectation 
value of $B$ is rather small. 
As a result, we observe a strong first-order transition 
from the gluonic phase to the NQ phase at $T \simeq 14~{\rm MeV}$. 
Note that $m_M^2$ remains positive at any value of $T$, which is 
consistent with the result shown in Fig. \ref{Figure1}. 

In Fig. \ref{Figure4}(b), the same plot is displayed 
for $\Delta_0=85~{\rm MeV}$. 
At low temperature, like in the case of $\Delta_0=75~{\rm MeV}$, 
the gluonic phase is more favored than the chromomagnetically stable NQ phase. 
At $T \simeq 9~{\rm MeV}$, $m_M^2$ turns negative, 
meaning that the stable NQ phase undergoes a phase transition 
into the unstable g2SC phase (see Fig. \ref{Figure1}). 
The gluonic phase is energetically favored 
until the temperature reaches $T \simeq 20~{\rm MeV}$. 
Above this temperature, the g2SC phase becomes stable 
and therefore is favored. 

In Figs. \ref{Figure4}(c) and \ref{Figure4}(d), 
we plot the free energy for the cases of $\Delta_0=100~{\rm MeV}$ and 
$\Delta_0=140~{\rm MeV}$, respectively. 
In both cases, the gluonic phase is energetically favored at low temperature. 
In contrast, at high temperature, 
the global minima of the free energy are realized at $B=0$, 
which means that, as expected from the result of Fig. \ref{Figure1}, 
the chromomagnetically stable 2SC/g2SC phases are favored. 
For $\Delta_0=100~{\rm MeV}$, the phase transition from the gluonic phase 
to the g2SC phase takes place at $T \simeq 21~{\rm MeV}$. 
In the case of $\Delta_0=140~{\rm MeV}$, the phase transition 
takes place at $T \simeq 18~{\rm MeV}$.

Here, we would like to make a comment regarding the order 
of the phase transitions. 
As mentioned above, we observed the strong first-order transition 
(gluonic phase $\leftrightarrow$ NQ phase) at $\Delta_0=75~{\rm MeV}$. 
On the other hand, for the cases of 
$\Delta_0=85,~100,~140~{\rm MeV}$, 
the phase transition (gluonic phase $\leftrightarrow$ 2SC/g2SC phases) 
is likely to be of second order. 
However, evaluating the free energy self-consistently 
near the critical temperatures is not easy 
and hence we do not exclude the possibility 
of weak first-order transitions. 
Furthermore, it should be also mentioned that, 
because of the cutoff artifact in Eq. (\ref{eqn:ep2}), 
it might be impossible to distinguish 
a weak first-order transition from a second-order one.

\section{Summary, Conclusions, and Outlook}
\subsection{Phase diagram}
We studied the gluonic cylindrical phase II 
at nonzero temperature. 
Using the gauged NJL model and the one-loop approximation, 
we computed the free energy of the gluonic phase self-consistently 
and investigated the phase structure of the gluonic phase. 
Although we neglected the color chemical potentials, 
we have checked that, for $\alpha_s \simeq 1$, 
their effect on the free energy is negligible.

\begin{figure}
\includegraphics[width=0.45\textwidth]{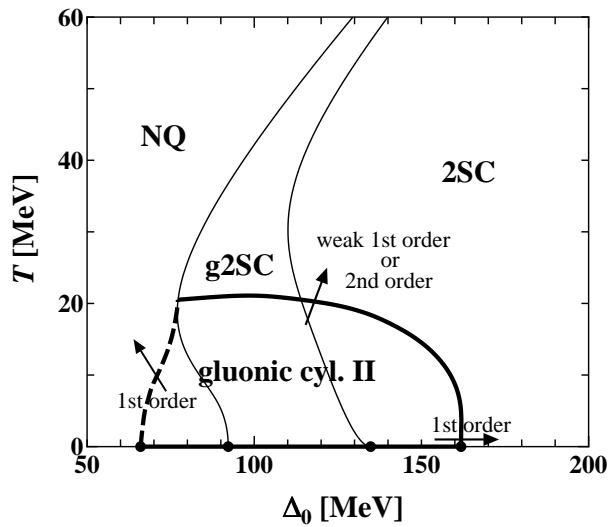}
\caption{Schematic phase diagram of neutral two-flavor quark matter 
at moderate density in $\Delta_0$-$T$ plane. 
The thick solid line denotes the line of 
second-order or weakly first-order transitions 
and strong first-order transitions are indicated by a thick dashed line. 
In the region enclosed by the thick solid and dashed lines, 
the gluonic phase is energetically more favored 
than the 2SC/g2SC/NQ phases.}
\label{Figure5}
\end{figure}

In the weak coupling regime, we found that the gluonic phase 
undergoes a strong first-order transition into the NQ phase as it is heated. 
This is a new aspect of the gluonic phase at $T>0$, which is not shown 
in Fig. \ref{Figure1}. 
On the other hand, since the phase transitions from the gluonic phase 
to the chromomagnetically stable 2SC/g2SC phases 
are of second-order or weakly first-order, 
we expect that the corresponding critical line shown in Fig. \ref{Figure1} 
(i.e., the right branch of the thick solid line) 
is not drastically altered by the self-consistent analysis. 
(In other words, the Meissner masses squared 
can be a rough criterion for choosing the energetically favored phase 
in this regime.) 
We thus are able to make a sketch of a schematic phase diagram 
of two-flavor quark matter, which is free from 
the chromomagnetic instability related to gluons 4--7 
(see Fig. \ref{Figure5}).
\footnote{It is interesting to note that the neutral single plane-wave 
Larkin-Ovchinnikov-Fulde-Ferrell state \cite{LOFF1,LOFF2} 
has a similar phase structure 
as the gluonic phase \cite{Gorbar2005b,Kiri2006,Kiri2006b,HJZ}.} 
The low-temperature region of the g2SC phase 
and a part of the 2SC phase is replaced by the gluonic phase. 
Furthermore, the gluonic phase wins against 
a part of the NQ phase and enlarges its region. 
We argue therefore that the gluonic phase which 
could resolve the chromomagnetic instability related to gluons 4--7 
is a strong candidate for the ground state 
of a neutral two-flavor color superconductor 
in the intermediate coupling regime. 
Alternatives include other types of the gluonic phases 
\cite{Gorbar2005,Gorbar2005a,Hashimoto2007,Ferrer}, the crystalline phases 
\cite{ABR,BR,Rajagopal2006} and the mixed phase \cite{RedRup}. 
It should be mentioned that, at $T=0$, 
the gluonic color-spin locked phase is more stable 
than the gluonic cylindrical phase II 
in some region of $\Delta_0$ 
and moreover is free from the chromomagnetic instability 
at moderate densities \cite{HashimotoJia,Hashimoto2008}.

Although we concentrated on the phase diagram 
in $T$-$\Delta_0$ plane in this work, 
it is obviously worthwhile to revisit the phase diagram 
in $T$-$\mu$ plane. A preliminary study \cite{Kiri2008}
indicates that currently known phase diagrams 
\cite{pd1,pd2,pd3} must be significantly altered. 
In addition, the critical temperature for the gluonic phase 
could reach a few tens of MeV and, therefore, 
it is interesting to study astrophysical implications 
of the gluonic phase, 
e.g., the quark matter equation of state, 
neutrino emission from compact star cores, and so on.

\subsection{Gluonic phase versus mixed phase}
Finally we briefly look at a mixed phase 
consisting of the NQ and the 2SC phases \cite{RedRup,Neumann,SHH}. 
For the mixed phase to exist, 
it must satisfy the Gibbs conditions, 
which are equivalent to chemical and mechanical equilibrium conditions 
between the NQ and the 2SC phases. 
These conditions end up as follows
\begin{eqnarray}
P^{{\rm (NQ)}}(\mu,\mu_e)=P^{{\rm (2SC)}}(\mu,\mu_e).\label{eqn:pbr}
\end{eqnarray}
Beside Eq. (\ref{eqn:pbr}) two components must 
have opposite electrical charge densities. 
Otherwise a globally neutral mixed phase could not exist. 
We solved Eq. (\ref{eqn:pbr}) and found 
that the globally neutral mixed phase exists in the window
\begin{eqnarray}
67~{\rm MeV} < \Delta_0 < 201~{\rm MeV},
\end{eqnarray}
where the quark chemical potential was taken to be 
$\mu=400~{\rm MeV}$.

\begin{figure}
\includegraphics[width=0.8\textwidth]{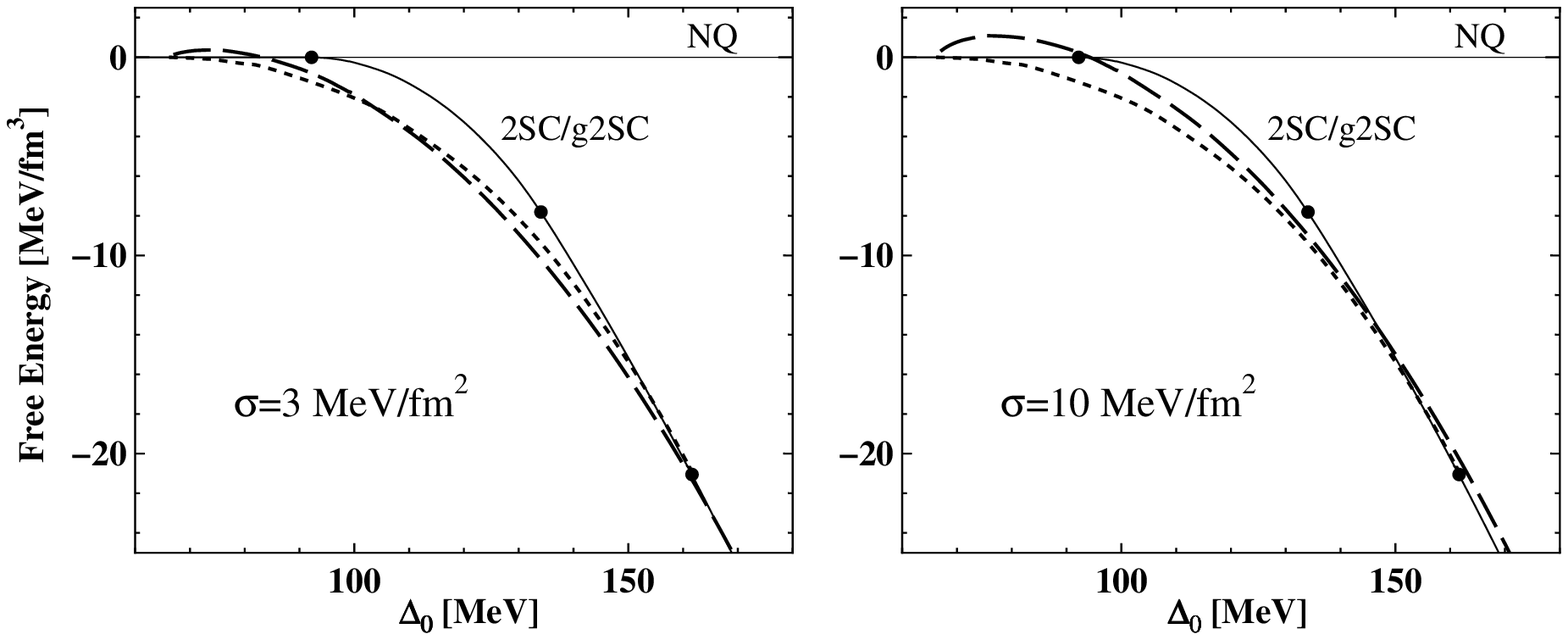}
\caption{The free energy of the neutral 2SC/g2SC phase 
(solid line), the gluonic phase (dotted line), 
and the mixed phase (dashed line) 
measured with respect to the NQ phase 
as a function of $\Delta_0$ for $\sigma=3~{\rm MeV/fm}^2$ (left) 
and $\sigma=10~{\rm MeV/fm}^2$ (right). 
The three dots on the solid line ($\Delta_0 = 92, 134, 162$ MeV 
from left to right) denote the edge of the g2SC window 
with the normal phase, the phase transition point 
between the 2SC and the g2SC phases, 
and the critical point of the chromomagnetic instability. 
The quark chemical potential is taken to be $\mu=400$ MeV.}
\label{Figure6}
\end{figure}

In order to calculate the free energy of the mixed phase 
we take account of finite-size effects, i.e., 
the surface and Coulomb energies 
associated with phase separation. 
The surface and Coulomb energy densities are given by
\begin{eqnarray}
\epsilon_S=\frac{dx\sigma}{r_0},~\epsilon_C=2\pi\alpha_{{\rm em}}
f_d(x)x(\varDelta n_e)^2r_0^2,
\end{eqnarray}
where $\sigma$ is the surface tension, $x$ is the volume fraction 
of the rarer phase, $\varDelta n_e$ is the difference 
of the electric charge density between NQ and 2SC phases, 
and $\alpha_{{\rm em}}=1/137$. 
These energy densities also depend on the dimension $d$ 
($d=1,2$, and 3 correspond to slabs, rods, 
and droplets configurations, respectively) 
and $r_0$, which denotes the radius of the rarer phase. 
The geometrical factor $f_d(x)$ is given by  
\begin{eqnarray}
f_d(x)=\frac{1}{d+2}\left(\frac{2-dx^{1-2/d}}{d-2}+x\right).
\end{eqnarray}
Minimizing the sum of $\epsilon_S$ and $\epsilon_C$ 
with respect to $r_0$, 
we obtain 
\begin{eqnarray}
\epsilon_S+\epsilon_C
=\frac{3}{2}\left(4\pi\alpha_{{\rm em}}d^2f_d(x)x^2\right)^{1/3}
(\varDelta n_e)^{2/3}\sigma^{2/3}.
\end{eqnarray}
The actual value of the surface tension 
in quark matter is poorly known, in this work 
we assume $d=3$ (droplets configuration) and 
try relatively small surface tension. 

Figure \ref{Figure6} displays the free energy 
of the 2SC/g2SC phase, the gluonic phase, and the mixed phase. 
For a very small surface tension $\sigma=3~{\rm MeV/fm}^2$, 
the mixed phase is the most favored in a wide range of $\Delta_0$, 
$103~{\rm MeV} < \Delta_0 < 166~{\rm MeV}$. 
The gluonic phase is energetically more favored 
than the mixed phase only in the weak coupling regime. 
Note that the value of the surface tension, 
$\sigma=3~{\rm MeV/fm}^2$ at $\mu=400~{\rm MeV}$, 
is close to that calculated by Reddy and Rupak \cite{RedRup}. 
For a surface tension $\sigma=10~{\rm MeV/fm}^2$, 
there still is a wide window where the mixed phase 
is more stable than the g2SC phase, 
but the mixed phase is less favored than the gluonic phase. 

It should be mentioned here that, however, 
we did not take into account the thickness of the boundary layer, 
which has been estimated to be comparable to the value of the 
Debye screening length in each of the two phases, 
and therefore the results shown in Fig. \ref{Figure6} is not 
a final conclusion \cite{Heiselberg,Gledenning}. 
The effect of charge screening would 
increase the surface energy substantially \cite{Norsen,Vorkre}.

\begin{acknowledgments}
I would like to thank Dirk Rischke and Armin Sedrakian 
for discussions and for comments on the earlier version of the manuscript. 
I also would like to thank H. Abuki, M. Ruggieri, and I. Shovkovy 
for discussions during the YITP international symposium 
``Fundamental Problems in Hot and/or Dense QCD''. 
This work was supported by the Deutsche Forschungsgemeinschaft (DFG).
\end{acknowledgments}

\end{document}